\documentstyle[prd,aps,eqsecnum,twocolumn,floats,psfig]{revtex}

\def\beq{\begin{equation}}
\def\eeq{\end{equation}}
\def\beqa{\begin{eqnarray}}
\def\eeqa{\end{eqnarray}}
\def\bfig{\begin{figure}}
\def\efig{\end{figure}}

\begin{document}
\draft
\fnsymbol{footnote}

\wideabs{

\title{Second-order rotational effects on the {\it r\,}-modes of
neutron stars}

\author{Lee Lindblom${}^1$, Gregory Mendell${}^2$, 
and Benjamin J. Owen${}^{1,3}$}

\address{${}^1$Theoretical Astrophysics 130-33,
         California Institute of Technology,
         Pasadena, CA 91125}
\address{${}^2$Department of Physics and Astronomy,
         University of Wyoming, 
         Laramie, WY 82071}
\address{${}^3$Max Planck Institut f\"ur Gravitationsphysik,
         Schlaatzweg 1, 14473 Potsdam, Germany}

\date{\today}
\maketitle
\begin{abstract}
  Techniques are developed here for evaluating the $r$-modes of
  rotating neutron stars through second order in the angular velocity
  of the star.  Second-order corrections to the frequencies and
  eigenfunctions for these modes are evaluated for neutron star
  models.  The second-order eigenfunctions for these modes are
  determined here by solving an unusual inhomogeneous hyperbolic
  boundary-value problem.  The numerical techniques developed to solve
  this unusual problem are somewhat non-standard and may well be of
  interest beyond the particular application here. The bulk-viscosity
  coupling to the $r$-modes, which appears first at second order, is
  evaluated.  The bulk-viscosity timescales are found here to be
  longer than previous estimates for normal neutron stars, but shorter
  than previous estimates for strange stars. These new timescales do
  not substantially affect the current picture of the gravitational
  radiation driven instability of the $r$-modes either for neutron
  stars or for strange stars.

\end{abstract}
}

\narrowtext

\section{Introduction}
\label{section0}

Recently the $r$-modes have been found to play an interesting and
important role in the evolution of hot young rapidly rotating neutron
stars.  Andersson~\cite{nils1} first realized and Friedman and
Morsink~\cite{friedman.morsink} confirmed more generally that
gravitational radiation tends to drive the $r$-modes unstable in all
rotating stars.  Lindblom, Owen, and Morsink~\cite{lom} then showed
that the coupling of gravitational radiation to the $r$-modes is
sufficiently strong to overcome internal fluid dissipation effects and
so drive these modes unstable in hot young neutron stars.  This result
has been verified by Andersson, Kokkotas, and Schutz~\cite{a.k.s}.
This result seemed somewhat surprising at first because the dominant
coupling of gravitational radiation to the $r$-modes is through the
current multipoles rather than the more familiar and usually dominant
mass multipoles.  But it is now generally accepted that gravitational
radiation does drive unstable any hot young neutron star with angular
velocity greater than about 5\% of the maximum (the angular velocity
where mass shedding occurs).  This instability therefore provides a
natural explanation for the lack of observed very fast pulsars
associated with young supernovae remnants.

The $r$-mode instability is also interesting as a possible source of
gravitational radiation.  In the first few minutes after the formation
of a hot young rapidly rotating neutron star in a supernova,
gravitational radiation will increase the amplitude of the $r$-mode
(with spherical harmonic index $m=2$) to levels where non-linear
hydrodynamic effects become important in determining its subsequent
evolution.  While the non-linear evolution of these modes is not well
understood as yet, Owen {\it et al.}~\cite{owen} have developed a
simple non-linear evolution model to describe it approximately.  This
model predicts that within about one year the neutron star spins down
(and cools down) to an angular velocity (and temperature) low enough
that the instability is again suppressed by internal fluid
dissipation.  All of the excess angular momentum of the neutron star
is radiated away via gravitational radiation.  Owen {\it et
al.}~\cite{owen} estimate the detectability of the gravitational waves
emitted during this spindown, and find that neutron stars spinning
down in this manner may be detectable by the second-generation
(``enhanced'') LIGO interferometers out to the Virgo cluster.
Bildsten~\cite{lars} and Andersson, Kokkotas, and
Stergioulas~\cite{aks2} have raised the possibility that the $r$-mode
instability may also operate in older colder neutron stars spun up by
accretion in low-mass x-ray binaries.  The gravitational waves emitted
by some of these systems (e.g.\ Sco X-1) may also be detectable by
enhanced LIGO~\cite{prb-tc}.  Thus, the $r$-modes of rapidly rotating
neutron stars have become a topic of considerable interest in
relativistic astrophysics.

The purpose of this paper is to explore further the properties of the
$r$-modes of rotating neutron stars.  The initial analyses of the
$r$-mode instability~\cite{nils1,friedman.morsink,lom} were based on a
small angular-velocity expansion for these modes developed originally
by Papaloizou and Pringle~\cite{p&p}.  This expansion in powers of the
angular velocity kept only the lowest-order terms in the expressions
for the various quantities associated with the mode: the frequency,
velocity perturbation, etc.  This lowest-order expansion is sufficient
to explore many of the interesting physical properties of these modes,
including the gravitational radiation instability.  However, some
important physical quantities vanish at lowest order and hence a
second-order analysis is needed~\cite{note0}.  For example the
coupling of the $r$-modes to bulk viscosity vanishes in the
lowest-order expansion.  Estimates of this important bulk-viscosity
coupling to the $r$-modes have been given by Lindblom, Owen, and
Morsink~\cite{lom}, Andersson, Kokkotas, and
Schutz~\cite{a.k.s,note0000}, and by Kokkotas, and
Stergioulas~\cite{KS}.  But none of these is based on the fully self
consistent second-order calculation needed to evaluate this coupling
accurately.  Since bulk viscosity is expected to be the dominant
internal fluid dissipation mechanism in hot young neutron stars, it is
important to extend the analysis so that this important physical
effect can be evaluated properly.

The dominant internal fluid dissipation mechanism in neutron stars
colder than about $10^9$K is thought to be a superfluid effect called
mutual friction~\cite{mendell-lind} caused by the scattering of
electrons off the magnetic fields in the cores of vortices.
Levin~\cite{levin} has shown that the importance of the $r$-mode
instability in low-mass x-ray binaries depends crucially on the
details of the mutual friction damping of these modes.  Unfortunately
the mutual friction dissipation also vanishes at lowest order in a
small angular velocity expansion of the superfluid $r$-modes.  Thus in
order to evaluate this effect properly, it is also necessary to
determine the structure of the $r$-modes of superfluid neutron stars
through second order in the angular velocity.  This provides another
motivation then for developing the tools needed to evaluate the
second-order rotational effects in the $r$-modes.

In this paper we develop a new formalism for exploring the
higher-order rotational effects in the $r$-modes.  Our analysis is
based on the two-potential formalism~\cite{ipser-lind} in which all
physical properties of a mode of a rotating star are expressed in
terms of two scalar potentials: a hydrodynamic potential $\delta U$
and the gravitational potential $\delta\Phi$.  We define a small
angular velocity expansion for the $r$-modes in terms of these
potentials, and derive the equations explicitly for the second-order
terms.  This expansion provides a straightforward and relatively
simple way to determine the second-order effects, such as the bulk
viscosity coupling, that are of interest to us here.  The equations
that determine the second-order terms in the $r$-modes form an
inhomogeneous hyperbolic boundary value problem that is not amenable
to solution by standard numerical techniques.  Therefore we have
developed new numerical techniques which could well have applications
beyond the present problem.  In particular these techniques will also
be needed to solve the analogous superfluid pulsation equations that
determine the effects of mutual friction on these modes.

The timescales derived here for the bulk viscosity damping of the
$r$-modes differ considerably from earlier estimates.  We find the
bulk-viscosity coupling to these modes to be weaker for normal neutron
stars than any previous estimates.  Consequently the gravitational
radiation-driven instability is somewhat more effective at driving
unstable the $r$-modes in hot young neutron stars than earlier
estimates suggested.  Although quantitatively different from earlier
estimates, our new values for the bulk-viscosity damping time do not
substantially alter the expected spindown scenario in hot young
neutron stars.  We re-evaluate the critical angular velocity curve
(above which the $r$-mode instability sets in) and find no qualitative
change from earlier estimates.  Our new value for the minimum critical
angular velocity is somewhat lower than earlier estimates: about 5\%
compared to about 8\% of the maximum.  In very hot young neutron stars
there is the possibility that bulk viscosity could re-heat the neutron
star (due in part to non-linear effects in the bulk viscosity) and so
suppress the instability to some extent~\cite{yuri}.  This could
result in a significant increase in the timescale required to spin
down young neutron stars, and could therefore decrease significantly
the detectability of the gravitational radiation emitted.  Our new
calculation of the bulk viscosity timescale indicates that reheating
will not be a major factor in the evolution of young neutron stars.
Our calculations also show that the bulk-viscosity coupling in strange
stars is somewhat stronger than the initial estimates by
Madsen~\cite{madsen2}.  We find that bulk viscosity completely
suppresses the $r$-mode instability in strange stars hotter than
$T\gtrsim 5\times 10^8$K, in good qualitative agreement with Madsen.

In Sec.~\ref{sectionI} we review the structure of equilibrium stellar
models through second order in the angular velocity of the star.  In
Sec.~\ref{sectionII} we review the two-potential formalism for
describing the modes of rotating stars, and derive the small angular
velocity expansion of these equations through second order.  In
Sec.~\ref{sectionIII} we focus our attention on the ``classical''
$r$-modes, the modes found previously to be subject to the
gravitational radiation driven instability.  We obtain analytical
expressions for the second-order corrections to the frequencies of
these modes, and present numerical results for polytropes and for more
realistic neutron star models.  In Sec.~\ref{sectionIV} we develop the
numerical techniques needed to find the second-order eigenfunctions
for the $r$-modes; we use those techniques to find those
eigenfunctions; and we present the results graphically. In
Sec.~\ref{sectionV} we use our new second-order expressions for the
$r$-modes to compute the effects of bulk viscosity on the evolution of
these modes. In the Appendix we discuss the convergence of the
numerical relaxation technique used in Sec.~\ref{sectionIV} to solve
the unusual hyperbolic boundary value problem for the second-order
eigenfunctions.

\section{Slowly Rotating Stellar Models}
\label{sectionI}

Our analysis of the $r$-modes of rotating stars is based on expanding
the equations as power series in the angular velocity $\Omega$ of the
star.  The first step therefore in obtaining these equations is to
find the structures of equilibrium stellar models in a similar power
series expansion.  This section describes how to solve the equilibrium
structure equations for uniformly rotating barotropic stars in such a
slow rotation expansion.  The solutions will be obtained here up to
and including the terms of order $\Omega^2$.

Let $h(p)$ denote the thermodynamic enthalpy of the barotropic fluid:

\beq
h(p) = \int_0^p {dp'\over \rho(p')},
\label{1.1}
\eeq

\noindent where $p$ is the pressure and $\rho$ is the density of the
fluid.  This definition can always be inverted to determine $p(h)$.
The barotropic equation of state, $\rho=\rho(p)$, then determines
$\rho(h)=\rho[p(h)]$.  The equations which determine the family of
stationary, axisymmetric uniformly rotating barotropic stellar models
are Euler's equation, which for this case has the simple form

\beq
0=\nabla_a\bigl[h-\case{1}{2}r^2(1-\mu^2)\Omega^2-\Phi\bigr],\label{1.2}
\eeq

\noindent and the gravitational potential equation,

\beq
\nabla^a\nabla_a\Phi =  -4\pi G\rho.\label{1.3}
\eeq

\noindent In these expressions $r$ and $\mu=\cos\theta$ are the
standard spherical coordinates, and $\Phi$ is the gravitational
potential.

We seek solutions to Eqs.~(\ref{1.2}) and (\ref{1.3}) as power series
in the angular velocity $\Omega$.  To that end, we define

\beq
h(r,\mu) = h_0(r) + h_2(r,\mu) {\Omega^2\over \pi G \bar{\rho}_0}
+ {\cal O}(\Omega^4),
\label{1.4}
\eeq

\beq
\rho(r,\mu) = \rho_0(r) + \rho_2(r,\mu) {\Omega^2\over \pi G \bar{\rho}_0}
+ {\cal O}(\Omega^4),
\label{1.5}
\eeq

\beq
\Phi(r,\mu) = \Phi_0(r) + \Phi_2(r,\mu) {\Omega^2\over \pi G \bar{\rho}_0}
+ {\cal O}(\Omega^4),
\label{1.6}
\eeq

\noindent where $\bar{\rho}_0$ is the average density of the
non-rotating star in the family.  Using these expressions then, the
first two terms in the solution to Eq.~(\ref{1.2}) are given by

\beq
C_0 = h_0(r) -\Phi_0(r),\label{1.7}
\eeq

\beq
C_2 = h_2(r,\mu) - \case{1}{2}\pi G \bar{\rho}_0 r^2(1-\mu^2) 
-\Phi_2(r,\mu),\label{1.8}
\eeq

\noindent where $C_0$ and $C_2$ are constants.  The non-rotating model
can be determined in the usual way by solving the gravitational
potential equation,

\beq
{1\over r^2} {d\over dr}\left(r^2{d\Phi_0\over dr}\right) = -
4\pi G \rho_0,\label{1.9}
\eeq

\noindent together with Eq.~(\ref{1.7}).  The integration constant,
$C_0$, can be shown to be $C_0=-GM_0/R_0$ by evaluating Eq.~(\ref{1.7})
at the surface of the star.  The constants $M_0$ and $R_0$ are the
mass and radius of the non-rotating star.

The second-order contributions to the stellar structure are determined
by solving the gravitational potential Eq.~(\ref{1.3}) together with
Eq.~(\ref{1.8}).  The second-order density perturbation $\rho_2$
is related to $h_2$ by

\beq
\rho_2(r,\mu) = \left({d\rho\over dh}\right)_0 h_2(r,\mu).
\label{1.10}
\eeq

\noindent Thus using Eq.~(\ref{1.8}), the equation for the second-order
gravitational potential can be written in the form

\beqa
\nabla^a&&\nabla_a \Phi_2+4\pi G \left({d\rho\over dh}\right)_0\Phi_2=
\nonumber \\
&&-4\pi G \left({d\rho\over dh}\right)_0\left\{C_2 
+ \case{1}{3}\pi G \bar{\rho}_0 r^2 [1-P_2(\mu)]
\right\},\label{1.11}
\eeqa

\noindent where $P_2(\mu)=\case{1}{2}(3\mu^2-1)$.  We note that the
right side of Eq.~(\ref{1.11}) splits into a function depending only on
$r$ plus a function of $r$ multiplied by $P_2(\mu)$.  Since the operator
on the left side of Eq.~(\ref{1.11}) acting on $P_2(\mu)$ gives a
function of $r$ multiplied by $P_2(\mu)$, it follows that the
second-order gravitational potential $\Phi_2$ must have a similar
splitting:

\beq
\Phi_2(r,\mu)=\Phi_{20}(r) + \Phi_{22}(r)P_2(\mu).\label{1.12}
\eeq

\noindent  Thus the partial differential equation (\ref{1.11})
for $\Phi_2$ reduces to a pair of ordinary differential equations
for the potentials $\Phi_{20}$ and $\Phi_{22}$:

\beqa
{1\over r^2}{d\over dr}\left(r^2 {d\Phi_{20}\over dr}\right)
&&+4\pi G \left({d\rho\over dh}\right)_0\Phi_{20}=\nonumber \\
&&-4\pi G \left({d\rho\over dh}\right)_0\left(C_2 + 
\case{1}{3} r^2 \pi G \bar{\rho}_0\right),\label{1.13}
\eeqa

\beqa
{1\over r^2}{d\over dr}\left(r^2 {d\Phi_{22}\over dr}\right)
&&-{6\over r^2}\Phi_{22}
+4\pi G \left({d\rho\over dh}\right)_0\Phi_{22}=\nonumber \\
&&\qquad\qquad\,\case{4}{3}\pi^2 G^2\bar{\rho}_0 
r^2 \left({d\rho\over dh}\right)_0.\label{1.14}
\eeqa

Appropriate boundary conditions are needed to select the unique
physically relevant solutions to Eqs.~(\ref{1.13}) and (\ref{1.14}).
In order to insure that the gravitational potential is non-singular
at the center of the star, $r=0$, we must require that $\Phi_{20}$
and $\Phi_{22}$ satisfy the following boundary conditions there:

\beq
0=\left({d\Phi_{20}\over dr}\right)_{r=0} = \Phi_{22}(0)
\label{1.15}
\eeq

\noindent The potential $\Phi_{22}$ must also fall to zero as
$r\rightarrow\infty$.  We can insure this by requiring that
$\Phi_{22}$ match smoothly at the surface of the star to a
potential that in the exterior of the star is proportional to
$P_2(\mu)/r^3$.  It is sufficient therefore to require that
$\Phi_{22}$ satisfy the condition

\beq \left({d\Phi_{22}\over dr}\right)_{r=R_0} = -{3\Phi_{22}(R_0)\over
R_0}.\label{1.16} \eeq

\noindent An additional condition is also needed to fix $\Phi_{20}$.  It
is customary to consider families of rotating stars which have the same
total mass.  In this case the monopole part of the exterior
gravitational potential is the same for all members of the family.  To
ensure this, we must require that the potential $\Phi_{20}$ and its
derivative vanish on the surface of the star:

\beq
0 = \Phi_{20}(R_0) = \left({d\Phi_{20}\over dr}\right)_{r=R_0}.
\label{1.17}
\eeq

\noindent It might appear that Eqs.~(\ref{1.15}) and  (\ref{1.17}) now
over constrain the potential $\Phi_{20}$.  This would be the case, except
that the constant $C_2$ that appears on the right side of
Eq.~(\ref{1.13}) is still undetermined.  The boundary conditions
Eqs.~(\ref{1.15}) through (\ref{1.17}) are just sufficient, however, to fix
uniquely the potentials $\Phi_{20}$ and $\Phi_{22}$ together with the
integration constant $C_2$ as solutions to Eqs.~(\ref{1.13}) and
(\ref{1.14}).  We also note that these boundary conditions insure
that 

\beq
0 = \int_{-1}^1\int _0^{R_0} r^2 \rho_2(r,\mu) drd\mu .\label{1.18}
\eeq

In summary then, the thermodynamic functions $h(r,\mu)$ and $\rho(r,\mu)$
in slowly rotating barotropic stars are given by Eqs.~(\ref{1.4}) and
(\ref{1.5}), where $\rho_2(r,\mu)$ and $h_2(r,\mu)$ are given by

\beqa
\rho_2(r,\mu) &&= \left({d\rho\over dh}\right)_0 h_2(r,\mu) \nonumber\\
&&=\left({d\rho\over dh}\right)_0 
\Bigl\{C_2 + \Phi_{20}(r) 
+ \case{1}{3} \pi G \bar{\rho}_0 r^2 \nonumber \\
&&\qquad\qquad
+ \left[\Phi_{22}(r)-\case{1}{3}\pi G \bar{\rho}_0 r^2\right]
P_2(\mu)\Bigr\}.\label{1.21}
\eeqa

\noindent These expressions for $h(r,\mu)$ and $\rho(r,\mu)$ depend
only on the structures of the non-rotating star through the functions
$h_0(r)$, $\rho_0(r)$, and $(d\rho/dh)_0$, the potentials $\Phi_{20}$
and $\Phi_{22}$ from Eqs.~(\ref{1.13}) and (\ref{1.14}), and the
constant $C_2$.

It is also instructive to work out an expression for the surface
$r=R(\mu,\Omega)$ of the rotating star.  This surface occurs where the
thermodynamic potential $h[R(\mu,\Omega),\mu]=0$.  Solving this
equation, we find

\beq
R(\mu,\Omega)= R_0 + R_2(\mu){\Omega^2\over \pi G \bar{\rho}_0}
+{\cal O}(\Omega^4),\label{1.22}
\eeq

\noindent where $R_2(\mu)$ is given by

\beqa
R_2(\mu) &&= R_{20} + R_{22}P_2(\mu)\nonumber\\
&&={3\over 4\pi G \bar{\rho}_0 R_0}
\Bigl\{C_2 + \case{1}{3}\pi G \bar{\rho}_0 R_0^2 \nonumber\\
&&\qquad
+ \bigl[\Phi_{22}(R_0) -\case{1}{3} \pi G \bar{\rho}_0 R_0^2\bigr]
P_2(\mu)\Bigr\}.
\label{1.23}
\eeqa

We have developed a computer code that solves these equations
numerically for stars with an arbitrary equation of state.  We have
tested this code against analytical expressions which can be obtained
for a polytropic neutron star equation of state, $p=K\rho^2$, with $K$
chosen so that a $1.4M_\odot$ model has a radius of $R_0=12.533$km.
We find that the constants that determine the slowly rotating model
for this polytropic case have the values $C_2=.09802 C_0$,
$R_{20}=.15198 R_0$, and $R_{22} = -.37995 R_0$.  Our numerical
results agree with the analytical to floating-point precision.

\section{The Pulsation Equations} \label{sectionII}

The modes of any uniformly rotating barotropic stellar model can be
described completely in terms of two scalar potentials $\delta U$ and
$\delta\Phi$~\cite{ipser-lind}.  The potential $\delta \Phi$ is the
Newtonian gravitational potential, while $\delta U$ determines the
hydrodynamic perturbation of the star:

\beq
\delta U = {\delta p\over \rho}-\delta\Phi,\label{2.1}
\eeq

\noindent where $\delta p$ is the Eulerian pressure perturbation, and
$\rho$ is the unperturbed density of the equilibrium stellar model.  We
assume here that the time dependence of the mode is $e^{i\omega t}$ and
that its azimuthal angular dependence is $e^{im\varphi}$, where $\omega$
is the frequency of the mode and $m$ is an integer.  The velocity
perturbation $\delta v^a$ is determined in this case by

\beq
\delta v^a = iQ^{ab}\nabla_b\delta U.\label{2.2}
\eeq

\noindent The tensor $Q^{ab}$ depends on the frequency of the mode,
and the angular velocity of the equilibrium star:

\beqa
Q^{ab}=&&{1\over (\omega+m\Omega)^2-4\Omega^2}\nonumber\\
&&\times\Biggl[(\omega+m\Omega)\delta^{ab}-
       {4\Omega^2\over \omega+m\Omega}z^az^b - 2i\nabla^av^b\Biggr].
\label{2.3}
\eeqa

\noindent In Eq.~(\ref{2.3}) the unit vector $z^a$ points along the
rotation axis of the equilibrium star, $\delta^{ab}$ is the
Euclidean metric tensor (the identity matrix in Cartesian coordinates), 
and $v^a$ is the velocity of the equilibrium stellar model.

In general, the potentials $\delta U$ and $\delta \Phi$ are solutions of
the following system of equations~\cite{ipser-lind}:

\beq \nabla_a(\rho Q^{ab}\nabla_b\delta U)=-(\omega+m\Omega)
{d\rho\over dh}(\delta U+\delta\Phi),\label{2.4} \eeq

\beq
\nabla^a\nabla_a\delta\Phi = -4\pi G{d\rho\over dh}
(\delta U +\delta\Phi),\label{2.5}
\eeq

\noindent subject to the appropriate boundary conditions at the
surface of the star for $\delta U$ and at infinity for $\delta \Phi$.
In order to discuss these boundary conditions in more detail we let
$\Sigma$ denote a function that vanishes on the surface of the star,
and which has been normalized so that its gradient,
$n_a=\nabla_a\Sigma$, is the outward directed unit normal vector
there, $n^an_a=1$.  The boundary condition on the function $\delta U$
at the surface of the star, $\Sigma=0$, is to require that the
Lagrangian perturbation in the enthalpy $h$ vanishes there, $\Delta h
=0$.  This condition can be written in terms of the variables used
here by noting that

\beq \Delta h = \delta h +\left({\delta v^a\over i\kappa
\Omega}\right)\nabla_ah,
\label{2.6}
\eeq

\noindent where $\kappa$ is related to the frequency of the mode by

\beq
\kappa\Omega = \omega+m\Omega.\label{2.7}
\eeq

\noindent Thus using Eqs.~(\ref{2.1}) and (\ref{2.2}) the boundary
condition can be written in terms of $\delta U$ and $\delta \Phi$ as

\beq
0=\biggl[\kappa\Omega(\delta U +\delta\Phi)
+ Q^{ab}\nabla_a h \nabla_b\delta U
\biggr]_{\Sigma\uparrow 0}.\label{2.8}
\eeq

\noindent The perturbed gravitational potential $\delta\Phi$ must vanish
at infinity, $\lim_{\,r\rightarrow \infty}\delta\Phi = 0$.  In addition
$\delta\Phi$ and its first derivative must be continuous at the surface
of the star.  The problem of finding the modes of uniformly rotating
barotropic stars is reduced therefore to finding the solutions to
Eqs.~(\ref{2.4}) and (\ref{2.5}) subject to the boundary condition in
Eq.~(\ref{2.8}).

The equation for the hydrodynamic potential $\delta U$, Eq.~(\ref{2.4}),
has a complicated dependence on the frequency of the mode and the
angular velocity of the star through $Q^{ab}$ as given in
Eq.~(\ref{2.3}).  In the analysis that follows it will be necessary to
have those dependences displayed more explicitly.  To that end, we
re-write Eq.~(\ref{2.4}) and the boundary condition Eq.~(\ref{2.8}) in
the following equivalent forms:

\beqa
\nabla^a\Bigl[&&\rho (\kappa^2\delta ^{ab}- 4z^az^b)\nabla_b\delta U\Bigr]
+ {2m\kappa\over \varpi}\varpi^a\nabla_a\rho\,\, \delta U \nonumber\\
&&\qquad\qquad\qquad=
-\kappa^2(\kappa^2-4)\Omega^2 {d\rho\over dh} (\delta U + \delta \Phi),
\label{2.9}
\eeqa

\beqa
\biggl[\bigl(\kappa^2\delta^{ab} &&- 4z^az^b\bigr)\nabla_ah\nabla_b\delta U
+{2m\kappa\over\varpi} \varpi^a\nabla_ah\,\,\delta U\nonumber\\
&&+\kappa^2(\kappa^2-4)\Omega^2(\delta U +\delta\Phi)
\biggr]_{\Sigma\uparrow 0}=0.\label{2.10}
\eeqa

\noindent Here we use the notation $\varpi$ for the cylindrical
radial coordinate, $\varpi = r\sqrt{1-\mu^2}$, and $\varpi^a$ to denote
the unit vector in the $\varpi$ direction.

Our purpose now will be to derive solutions to Eqs.~(\ref{2.5}) and
(\ref{2.9}) as power series in the angular velocity of the star.  To
that end we define the expansions of the potentials $\delta U$ and
$\delta \Phi$ as

\beq \delta U = R_0^2\Omega^2\left[\delta U_0 + \delta U_2 
{\Omega^2\over \pi G\bar{\rho}_0} + {\cal O}(\Omega^4)
\right],\label{2.11} \eeq

\beq
\delta \Phi = R_0^2\Omega^2\left[\delta \Phi_0 
+ \delta \Phi_2 {\Omega^2\over \pi G \bar{\rho}_0}
+ {\cal O}(\Omega^4)\right].\label{2.12}
\eeq

\noindent The normalizations of $\delta U$ and $\delta\Phi$ have been
chosen to make the $\delta U_i$ and $\delta \Phi_i$ dimensionless
under the assumption that the lowest order terms scale as $\Omega^2$.
Here we have limited our consideration to the generalized
$r$-modes~\cite{li99}: modes which are dominated by rotational effects
and whose frequencies vanish linearly therefore in the angular
velocity of the star.  In this case $\kappa$ (as defined in
Eq.~\ref{2.7}) is finite in the small angular velocity limit, and so
we may expand

\beq
\kappa=\kappa_0 + \kappa_2 
{\Omega^2\over \pi G \bar{\rho}_0}
+ {\cal O}(\Omega^4).\label{2.13}
\eeq

\noindent Using these expansions,
together with those for the structure of the equilibrium star from
Eqs.~(\ref{1.4}) and (\ref{1.5}), it is straightforward to write down
order by order the equations for the mode.  The lowest order terms in
the expansions of Eqs.~(\ref{2.9}) and (\ref{2.5}) are the following,

\beq \nabla^a\Bigl[\rho_0 (\kappa_0^2\delta ^{ab} -
4z^az^b)\nabla_b\delta U_0\Bigr] + {2m\kappa_0\over
\varpi}\varpi^a\nabla_a\rho_0\,\, \delta U_0 =0, \label{2.14} \eeq

\beq
\nabla^a\nabla_a\delta\Phi_0 = - 4\pi G \left({d\rho\over dh}\right)_0
(\delta U_0 + \delta \Phi_0),
\label{2.15}
\eeq

\noindent Similarly, the lowest order term in the expansion of the boundary
condition is

\beqa
\biggl[
\bigl(\kappa_0^2\delta^{ab} 
- &&4z^az^b\bigr)\nabla_ah_0\nabla_b\delta U_0\nonumber \\
&&+{2m\kappa_0\over\varpi} \varpi^a\nabla_ah_0\,\,\delta U_0
\biggr]_{r=R_0}=0.
\label{2.16}
\eeqa

 Continuing on to second order, the equations for the potentials are

\beqa
\nabla_a\Bigl[&&\rho_0(\kappa_0^2\delta^{ab}-4z^az^b)\nabla_b\delta U_2\Bigr]
+{2m\kappa_0\over \varpi}\varpi^a\nabla_a\rho_0\,\,\delta U_2\nonumber\\
&&+\nabla_a\Bigl[\rho_2(\kappa_0^2\delta^{ab}-4z^az^b)\nabla_b\delta U_0
+2\kappa_0\kappa_2\rho_0\nabla^a\delta U_0\Bigr]\nonumber \\
&&+{2m\over\varpi}\varpi^a\left(\kappa_2\nabla_a\rho_0
+\kappa_0\nabla_a\rho_2\right)\delta U_0 =\nonumber \\
&&\qquad -\kappa_0^2(\kappa_0^2-4) \pi G \bar{\rho}_0
\left({d\rho\over dh}\right)_0(\delta U_0 +\delta\Phi_0),
\label{2.17}
\eeqa

\beqa
\nabla^a\nabla_a\delta\Phi_2 = &&-4\pi G \left({d\rho\over dh}\right)_0
(\delta U_2 + \delta \Phi_2)\nonumber \\
&&-4\pi G \left({d^2\rho\over dh^2}\right)_0h_2
(\delta U_0 + \delta \Phi_0).\label{2.18}
\eeqa

\noindent The second-order boundary condition is somewhat more
complicated; it must include two types of terms.  The first type comes
from the second-order terms in the expansion of Eq.~(\ref{2.8}) in
powers of the angular velocity.  The second type comes from the fact
that the boundary condition is to be imposed on the actual surface of
the rotating star, not the surface $r=R_0$.  This second type of term is
the correction to the lowest-order boundary condition, Eq.~(\ref{2.16}),
needed to impose it on the actual boundary of the star (to second order
in the angular velocity).  Hence the terms of the second type are
proportional to $R_2$, the second-order change in the radius of the star
from Eq.~(\ref{1.23}).  The resulting boundary condition is

\beqa
\Biggl\{(&&\kappa_0^2\delta^{ab} - 4z^az^b)\nabla_ah_0\nabla_b\delta U_2 +
{2m\kappa_0\over \varpi}\varpi^a\nabla_ah_0\delta U_2\nonumber \\
&&+(\kappa_0^2\delta^{ab} - 4z^az^b)\nabla_ah_2\nabla_b\delta U_0 +
{2m\kappa_0\over \varpi}\varpi^a\nabla_ah_2\delta U_0\nonumber \\
&&+2\kappa_0\kappa_2\nabla^ah_0\nabla_a\delta U_0
+{2m\kappa_2\over \varpi}\varpi^a\nabla_a h_0\delta U_0\nonumber \\
&&+\kappa_0^2(\kappa_0^2-4)\pi G\bar{\rho}_0(\delta U_0+\delta\Phi_0)
\nonumber \\
&&+R_2r^c\nabla_c\biggl[(\kappa_0^2\delta^{ab}-4z^az^b)
\nabla_ah_0\nabla_b\delta U_0
\nonumber \\
&&\qquad\qquad\qquad+{2m\kappa_0\over \varpi}
\delta U_0\varpi^a\nabla_ah_0\biggr]
\Biggr\}_{r=R_0}=0.
\label{2.19}
\eeqa

\noindent In summary then, Eqs.~(\ref{2.17}) and (\ref{2.18}) together
with the boundary condition Eq.~(\ref{2.19}) determine the
second-order terms in the structure of any generalized $r$-mode.

\section{The Classical $r$-Modes}
\label{sectionIII}

There exists a large class of modes in rotating barotropic stellar
models whose properties are determined primarily by the rotation of
the star~\cite{li99,fried-lockitch}.  We refer to these as generalized
$r$-modes.  In this section we restrict our attention however to those
modes which contribute primarily to the gravitational radiation driven
instability.  These ``classical'' $r$-modes (which were studied first by
Papaloizou and Pringle~\cite{p&p}) are generated by hydrodynamic
potentials of the form (see e.g. Lindblom and Ipser~\cite{li99})

\beq
\delta U_0 = \alpha \left({r\over R_0}\right)^{m+1} 
P_{m+1}^m(\mu)e^{im\varphi}.
\label{3.1}
\eeq

\noindent It is straightforward to verify that this $\delta U_0$ is a
solution to Eq.~(\ref{2.14}) if the eigenvalue $\kappa_0$ has the
value

\beq
\kappa_0 = {2\over m+1}.\label{3.2}
\eeq

\noindent This $\delta U_0$ and $\kappa_0$ also satisfy the boundary
condition Eq.~(\ref{2.16}) without further restriction at the boundary
(and at every point within the star as well).  The gravitational
potential $\delta \Phi_0$ must have the same angular dependence as
$\delta U_0$.  Thus, $\delta \Phi_0$ must (through a slight abuse of
notation) have the form

\beq
\delta\Phi_0 = \alpha \delta \Phi_0(r) P_{m+1}^m(\mu) e^{im\varphi}.
\label{3.3}
\eeq

\noindent The gravitational potential Eq.~(\ref{2.15}) reduces
to an ordinary differential equation then for $\delta\Phi_0(r)$:

\beqa
{d^2\delta\Phi_0\over dr^2}+{2\over r}{d\delta\Phi_0\over dr}
+ &&\biggl[4\pi G \left({d\rho\over dh}\right)_0 \!\!\!
- {(m+1)(m+2)\over r^2}
\biggr]\delta\Phi_0\nonumber \\
&& = -4\pi G \left({d\rho\over dh}\right)_0 \left({r\over R_0}\right)^{m+1}.
\label{3.4}
\eeqa

\noindent Once $\delta U_0$ and $\delta\Phi_0$ are known, it is
straightforward to evaluate the perturbations in other thermodynamic
quantities to this order.  For example $\delta p_0 = \rho_0 \delta h_0
= \rho_0 ( \delta U_0 +\delta \Phi_0)$.  And it is straightforward to
evaluate then the velocity perturbation to this order using
Eq.~(\ref{2.2}).

We next consider the second-order contributions to the $r$-modes.
First, let us analyze the second-order equation for the potential
$\delta U$, Eq.~(\ref{2.17}).  This equation contains two types of
terms: those proportional to $\delta U_2$, and those that are not.  We
will consider those terms not proportional to $\delta U_2$ as source
terms, and we evaluate them now.  It is convenient to break these
source terms into three groups.  The first group is proportional to
$\rho_2$. These terms can be simplified by recalling
that $\delta U_0$ satisfies Eq.~(\ref{2.14}) for {\it any} spherically
symmetric density distribution.  Then, using the fact from
Eq.~(\ref{1.21}) that $\rho_2(r,\mu)=\rho_{20}(r) +
\rho_{22}(r)P_2(\mu)$, we find

\beqa
\nabla_a\Bigl[\rho_2(\kappa_0^2\delta^{ab}&&-4z^az^b)\nabla_b\delta U_0\Bigr]
+ {2m\kappa_0\over \varpi}\varpi^a\nabla_a\rho_2\delta U_0=\nonumber\\
&&\qquad\qquad\qquad
 -{12m(m+2)\over (m+1)^2} {\rho_{22}\over r^2}\delta U_0.
\label{3.6}
\eeqa

\noindent The second group of terms is proportional to $\kappa_2$.
These terms have the following simplified form:

\beqa
\nabla^a\bigl(2\kappa_0\kappa_2\rho_0\nabla_a\delta U_0\bigr)
&&+ {2m\kappa_2\over \varpi} \varpi^a\nabla_a \rho_0\delta U_0 =
\nonumber \\
&& \qquad\qquad 2(m+2)\kappa_2{1\over r}{d\rho_0\over dr}\delta U_0.
\label{3.7}
\eeqa

\noindent Thus, combining together these terms with those on the right
side of Eq.~(\ref{2.17}), we obtain the following expression for
the equation that determines $\delta U_2$ for the classical $r$-modes:

\beqa
\nabla_a\biggl\{&&\rho_0\biggl[
{4\delta^{ab}\over(m+1)^2}-4z^az^b\biggr]\nabla_b\delta U_2\biggr\}
\!+\!{4m\varpi^a\nabla_a\rho_0\over(m+1)\varpi}\delta U_2=\nonumber\\
&&{12m(m+2)\over (m+1)^2}{\rho_{22}\over r^2} \delta U_0
-2(m+2)\kappa_2{1\over r}{d\rho_0\over dr} \delta U_0\nonumber \\
&&+16\pi G\bar{\rho}_0 {m(m+2)\over (m+1)^4}
\left({d\rho\over dh}\right)_0(\delta U_0+\delta \Phi_0).
\label{3.8}
\eeqa

A similar reduction can also be made on the second-order boundary
condition, Eq.~(2.19).  We collect similar terms together to obtain
the following simplifications:

\beqa
(\kappa_0^2\delta^{ab}&&-4z^az^b)\nabla_a h_2\nabla_b\delta U_0
+{2m\kappa_0\over\varpi}\varpi^a\nabla_a h_2\, \delta U_0 \nonumber \\
&&=-{12m(m+2)\over (m+1)^2} {h_{22}\over r^2} \delta U_0,
\label{3.9}
\eeqa

\beqa
R_2r^c\nabla_c\biggl[(\kappa_0^2\delta^{ab}&&-4z^az^b)
\nabla_ah_0\nabla_b\delta U_0
\nonumber \\
&&+{2m\kappa_0\over \varpi} \delta U_0\varpi^a
\nabla_ah_0\biggr]=0.
\label{3.10}
\eeqa

\noindent The latter follows from the fact that the expression in
Eq.~(\ref{2.16}) is zero everywhere if $\delta U_0$ is given by
Eq.~(\ref{3.1}).  Combining these simplified expressions together
gives the following form for the boundary condition that constrains
$\delta U_2$:

\beqa
\biggl\{\biggr[&&{4\delta^{ab}\over (m+1)^2} - 4z^az^b\biggr]
\nabla_ah_0\nabla_b\delta U_2 +
{4m\varpi^a\nabla_ah_0\over (m+1)\varpi}\delta U_2\nonumber \\
&&-{12m(m+2)\over (m+1)^2} {h_{22}\over r^2} \delta U_0
+2(m+2)\kappa_2{1\over r}{dh_0\over dr} \delta U_0\nonumber \\
&&-{16m(m+2)\over (m+1)^4}\pi G \bar{\rho}_0 (\delta U_0 +\delta \Phi_0)
\Biggr\}_{r=R_0}=0.\nonumber \\
\label{3.11}
\eeqa

We note that the operator on the left side of Eq.~(\ref{3.8}) which acts
on $\delta U_2$ is identical to the operator that acts on $\delta U_0$
from the lowest-order Eq.~(\ref{2.14}).  We also note that the right side
of Eq.~(\ref{3.8}) is a function of $r$ multiplied by the angular
function $P_{m+1}^m(\mu)e^{im\varphi}$.  These facts allow us to derive
a simple formula for the second-order eigenvalue $\kappa_2$ in terms of
known quantities.  Multiply the left side of Eq.~(\ref{3.8}) by $\delta
U_0^*$ and integrate over the interior of the star.  This integral
vanishes because this operator is symmetric and $\delta U_0^*$ also
satisfies Eq.~(\ref{2.14}).  This implies that the integral of $\delta
U_0^*$ multiplied by the right side of Eq.~(\ref{3.8}) also vanishes. 
This integral gives the following expression for the eigenvalue
$\kappa_2$ once the angular integrals are performed:

\beqa \kappa_2 &&\int_0^{R_0} \left({r\over
R_0}\right)^{2m+2}r{d\rho_0\over dr} dr = \nonumber \\
&&\quad{6m\over (m+1)^2}
\int_0^{R_0} \rho_{22}\left({r\over R_0}\right)^{2m+2} dr
\nonumber\\&&\quad+{8\pi G
\bar{\rho}_0 m\over(m+1)^4} \int_0^{R_0}
r^2\left({r\over R_0}\right)^{m+1}
\nonumber \\&&\quad\qquad
\times\biggl[\left({r\over R_0}\right)^{m+1}
+\delta\Phi_0(r)\biggr] \left({d\rho\over
dh}\right)_0 dr.\label{3.12}
\eeqa

\begin{table}
\caption{The second-order eigenvalues $\kappa_2$ of the classical
$r$-modes for stars with polytropic equations of state
$p=K\rho^{1+1/n}$.\label{table1}}
\begin{tabular}{cddddd}
$n$ &$m=2$ &$m=3$ &$m=4$ &$m=5$&$m=6$\\
\tableline
0.0 &.57407 &.59766 & .54720 & .49074 & .44044 \\
0.5 &.41718 &.43861 & .40415 & .36406 & .32782 \\
1.0 &.29883 &.32054 & .29946 & .27250 & .24729 \\
1.5 &.21183 &.23426 & .22369 & .20693 & .19019 \\
2.0 &.14777 &.17084 & .16846 & .15961 & .14942 \\
2.5 &.10091 &.12426 & .12808 & .12532 & .12016 \\
3.0 &.06716 &.09024 & .09859 & .10039 & .09905 \\
3.5 &.04334 &.06556 & .07699 & .08210 & .08357 \\
4.0 &.02692 &.04773 & .06102 & .06839 & .07186 \\
4.5 &.01589 &.03487 & .04896 & .05768 & .06252 \\
 \end{tabular}
\end{table}

 We have evaluated Eq.~(\ref{3.12}) numerically to determine
$\kappa_2$ for a variety of equations of state.  Table~\ref{table1}
presents the values of $\kappa_2$ for the classical $r$-modes with
$2\leq m \leq 6$ of stars with polytropic equations of state.  We also
present in Figure~\ref{fig1} a graph of the frequency
$\omega/\Omega=\kappa-2$ of the $m=2$ classical $r$-modes computed for
$1.4$M${}_\odot$ stellar models based on seven realistic equations of
state~\cite{salgado}.  The dashed line in Fig.~\ref{fig1} corresponds to the
lowest-order approximation of the $r$-mode frequency
$\omega/\Omega=\kappa_0-2$, which is the same for any equation of
state.  The solid curves are based on the second-order formula
$\omega/\Omega=\kappa_0-2+\kappa_2 \Omega^2/\pi G\bar{\rho}_0$.  It is
interesting to see in Fig.~\ref{fig1} that the higher-order terms make
only small (up to about $12\%$ at the highest angular velocities)
corrections to the frequencies of these modes for stars with realistic
equations of state.  We also note that the general tendency of the
frequency of these modes to be smaller than that predicted by the
lowest-order expression is consistent with the results found by
Lindblom and Ipser~\cite{li99} for the Maclaurin spheroids.  An
analytical expression for $\kappa_2$ can be obtained from
Eq.~(\ref{3.12}) for the uniform density case by performing the
indicated integrals analytically.  The resulting expression is
equivalent to Eq.~(6.10) of Lindblom and Ipser~\cite{li99}.

\bfig \centerline{\psfig{file=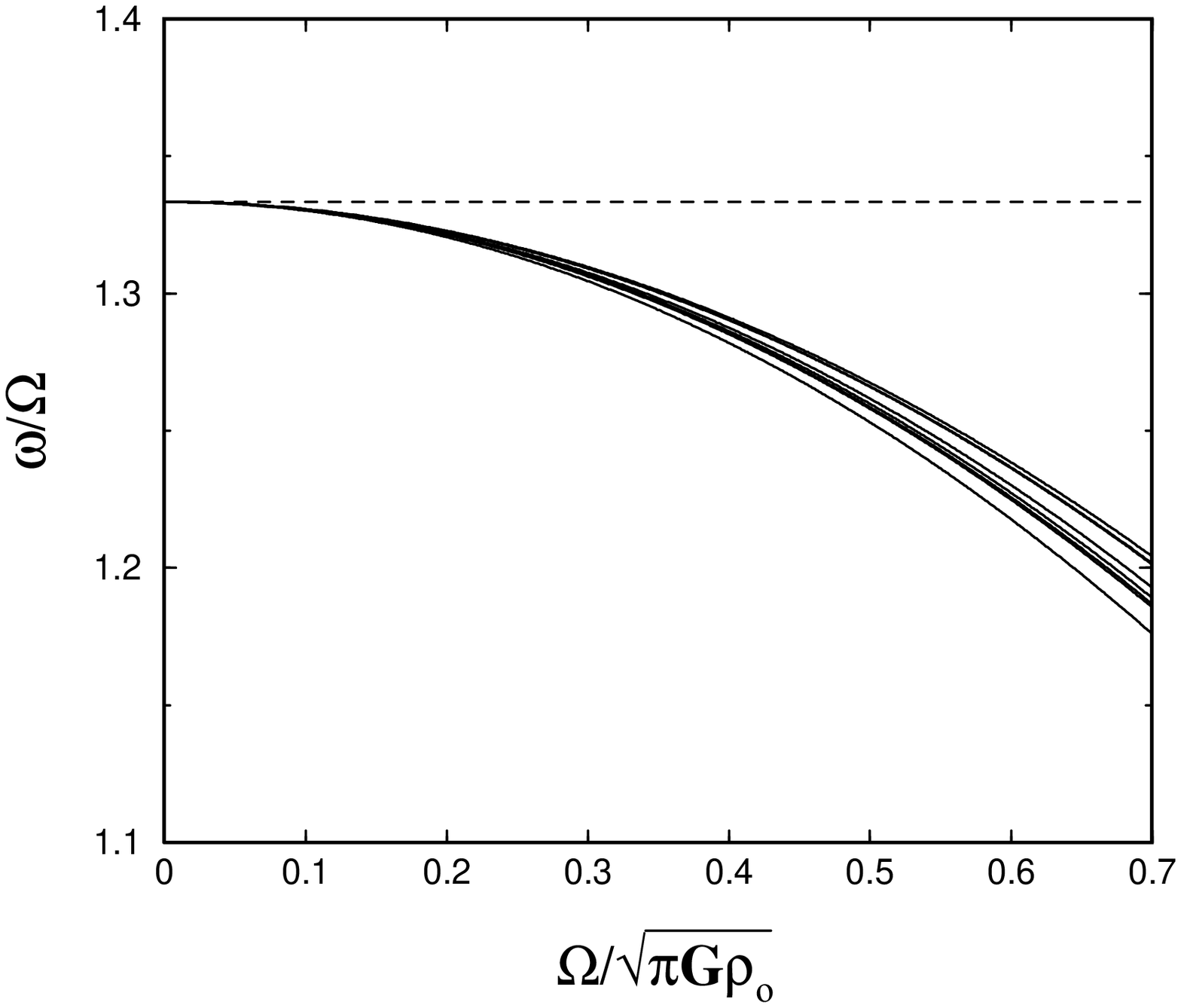,height=2.4in}} \vskip 0.3cm
\caption{Angular velocity dependence of the frequencies of the
classical $m=2$ $r$-modes for 1.4M${}_\odot$ stellar models based on
seven realistic neutron star equations of state.  The dashed curve is
based on the lowest-order expression for $\omega/\Omega=\kappa_0-2$,
while the solid curves are based on the second-order
expression $\omega/\Omega=\kappa_0-2+\kappa_2
\Omega^2/\pi G\bar{\rho}_0$.\label{fig1}} \efig

\section{Numerical Solutions for $\delta U_2$}
\label{sectionIV}

In this section we discuss the numerical solution of the equations
that determine the second-order corrections to the eigenfunctions
$\delta U_2$ and $\delta \Phi_2$ of the classical $r$-modes.  Once
$\delta U_2$ is known, the solution of Eq.~(\ref{2.18}) to determine
$\delta \Phi_2$ is straightforward.  Thus our discussion will
concentrate on the more difficult problem of solving Eq.~(\ref{3.8})
for $\delta U_2$.  It will be convenient to introduce the notation

\beqa
D(\delta U_2)=
\nabla_a\biggl\{\rho_0\biggl[{4\delta^{ab}\over(m+1)^2}&&-4z^az^b\biggr]
\nabla_b\delta U_2\biggr\}
\nonumber\\
&&+{4m\varpi^a\nabla_a\rho_0\over(m+1) \varpi}\delta U_2,
\label{4.1}
\eeqa

\noindent for the differential operator that appears on the left side
of Eq.~(\ref{3.8}).  Thus Eq.~(\ref{3.8}) can be written in the form:

\beq
D(\delta U_2) = F,\label{4.2}
\eeq

\noindent where 

\beqa
F =
&&{12m(m+2)\over (m+1)^2}{\rho_{22}\over r^2} \delta U_0
-2(m+2)\kappa_2{1\over r}{d\rho_0\over dr} \delta U_0\nonumber \\
&&+16\pi G\bar{\rho}_0 {m(m+2)\over (m+1)^4}
\left({d\rho\over dh}\right)_0(\delta U_0+\delta \Phi_0).
\label{4.3}
\eeqa

\noindent The problem of solving Eq.~(\ref{4.2}) numerically is a
somewhat non-standard problem that is made difficult by two facts.
First, the operator $D$ has a non-trivial kernel: $D(\delta U_0)=0$.
Many of the straightforward numerical techniques fail in this case.
Second, the operator $D$ is hyperbolic.  There appears to be little
previous work on solving hyperbolic boundary value problems of this
type.

We solve Eq.~(\ref{4.2}) here using a variation of the standard
relaxation method commonly used to solve elliptic partial differential
equations~\cite{recipes}.  To that end we introduce a fictitious time
parameter $\lambda$ and convert Eq.~(\ref{4.2}) into an evolution
equation

\beq
\partial_\lambda \delta U_2 = D(\delta U_2)-F.\label{4.4}
\eeq

\noindent The idea is to impose as initial data for Eq.~(\ref{4.4}) a
guess for $\delta U_2$, and then to evolve these data (as a function
of $\lambda$) until a stationary ($\partial_\lambda\delta U_2=0$)
state is reached.  If successful, the late time solution
($\lim_{\lambda \rightarrow\infty}\delta U_2$) to Eq.~(\ref{4.4}) will
also be a solution to Eq.~(\ref{4.2}).

We implement the relaxation method to solve Eq.~(\ref{4.2}) by using a
discrete representation of the functions and differential operators.
Let $u_n^i$ denote the discrete representation of the function $\delta
U_2$ evaluated at the fictitious time $\lambda_n$.  The index $i$ (and
later $j$ and $k$ as well) takes on values from $1$ to $N$ where $N$
is the dimension of the particular discretization used.  Similarly the
discrete representation of the differential operator $D$ of
Eq.~(\ref{4.2}) is denoted $D^j{}_i$, and the representation of the
right side of Eq.~(\ref{4.2}) is denoted $F^j$.  Thus the discrete
representation of Eq.~(\ref{4.2}) is simply

\beq
D^j{}_i u{}^i = F^j.\label{4.5}
\eeq

\noindent Summation is implied for pairs of repeated indices (e.g. $i$
on the left side of Eq.~\ref{4.5}).  The algebraic Eq.~(\ref{4.5})
cannot be solved by the most straightforward direct numerical
techniques because the operator $D$ has a nontrivial kernel (mentioned above).
Consequently the matrix $D^j{}_i$ has no inverse.

Thus we are lead to introduce the evolution Eq.~(\ref{4.4}).  We use
the ``implicit'' form of the discrete representation of Eq.~(\ref{4.4}):

\beq
O^j{}_iu_{n+1}^i
\equiv\left(D^j{}_i-{1\over\Delta\lambda}I^j{}_i \right)u_{n+1}^i 
= F^j - {1\over\Delta\lambda}u_n^j.\label{4.6}
\eeq

\noindent In Eq.~(\ref{4.6}) $I^j{}_i$ is the $N$ dimensional identity
matrix, and $\Delta\lambda$ is the relaxation timestep.  Given $u_n^i$
we solve Eq.~(\ref{4.6}) for $u_{n+1}^i$ by direct solution of the
linear algebraic equation.  We use the band-diagonal linear equation
solver LINSIS from EISPACK to compute
$(O^{-1})^i{}_j(F^j-u^j_n/\Delta\lambda)$.  We find that this can be
computed stably and accurately for almost any value of the relaxation
timestep $\Delta\lambda$. 

Unfortunately solving Eq.~(\ref{4.6}) iteratively does not yield the
desired solution to Eq.~(\ref{4.5}) in the limit of large $n$.
Instead the solution grows exponentially, becoming in the limit of
large $n$ closer and closer to a non-trivial solution to the
homogeneous equation, $\delta U_0$.  Fortunately, this malady is
easily corrected.  Let $\bar{u}^i$ denote the discrete representation
of $\delta U_0$; thus $D^j{}_i \bar{u}^i\approx 0$ since $\delta U_0$
is in the kernel of $D$.  Also we let $\bar{u}_i$ denote the discrete
representation of the co-vector associated with $\bar{u}^i$.  In
particular choose $\bar{u}_i$ so that $\bar{u}^i\bar{u}_i$ is the
discrete representation of the integral of $|\delta U_0|^2$.  Then
the matrix

\beq
P^j{}_i=I^j{}_i - {\bar{u}^j\bar{u}_i\over \bar{u}^k\bar{u}_k},
\label{4.7}
\eeq
\noindent is the discrete representation of the operator that projects
functions into the subspace orthogonal to $\delta U_0$.  We use this
projection in conjunction with Eq.~(\ref{4.6}) to define a modified
relaxation scheme to determine $u_{n+1}^i$ iteratively:

\beq
u_{n+1}^i = P^i{}_j(O^{-1}){}^j{}_k
\left(F^k - {u_n^k\over \Delta\lambda}\right).
\label{4.8}
\eeq

\noindent By applying the projection $P^i{}_j$ after each relaxation
step we insure that the exponentially growing kernel is removed from
the solution.  We find that the iteration scheme defined in
Eq.~(\ref{4.8}) does converge quickly and stably to a solution of
Eq.~(\ref{4.5}).  In the Appendix we discuss the reason this numerical
relaxation method works even in the case of the unusual hyperbolic
boundary-value problem considered here.  We show that convergence
is guaranteed for sufficiently large values of the relaxation timestep
$\Delta\lambda$, and that it also converges for either sign of $\Delta
\lambda$.

In order to implement this inversion scheme we need explicit discrete
representations of these operators.  We find it convenient to work in
spherical coordinates $r$ and $\mu=\cos\theta$.  In terms of these
coordinates then, the differential operator $D$ has the form
\vfill\eject

\beqa
&&D(\delta U_2)
={4\rho_0\over (m+1)^2} \biggl[ {\partial^2\delta U_2\over\partial r^2}
+{1-\mu^2\over r^2}{\partial^2\delta U_2\over \partial\mu^2}
+{2\over r}{\partial\delta U_2\over \partial r}\nonumber \\
&&\qquad\qquad\quad
-{2\mu\over r^2}{\partial\delta U_2\over\partial\mu}
-{m^2\delta U_2\over r^2(1-\mu^2)}\biggr]
-4\rho_0\biggl[\mu^2{\partial^2\delta U_2\over\partial r^2}
\nonumber \\
&&\qquad\qquad\quad
+{2\mu(1-\mu^2)\over r}{\partial^2 \delta U_2\over \partial r\partial\mu}
+{(1-\mu^2)^2\over r^2}{\partial^2\delta U_2\over \partial\mu^2}
\nonumber \\
&&\qquad\qquad\quad
+{1-\mu^2\over r}{\partial\delta U_2\over \partial r}
-{3\mu(1-\mu^2)\over r^2}{\partial\delta U_2\over \partial\mu}\biggr]
\nonumber \\
&&\qquad\quad
+{4\over (m+1)^2}\biggl({d\rho\over dh}\biggr)_0 {dh_0\over dr}\biggl[
[1-(m+1)^2\mu^2]{\partial\delta U_2\over \partial r}\nonumber\\
&&\qquad\qquad
-(m+1)^2{\mu(1-\mu^2)\over r}{\partial\delta U_2\over \partial\mu}
+m(m+1){\delta U_2\over r}\biggr].\nonumber\\
\label{4.9}
\eeqa

\noindent A similar spherical representation is also needed for the
boundary condition, Eq.~(3.11),

\beqa
\Biggl\{
\Bigl[1&&-(m+1)^2\mu^2\Bigr]{dh_0\over dr}{\partial \delta U_2\over \partial r}
+m(m+1){1\over r}{dh_0\over dr}\delta U_2\nonumber \\
&&-(m+1)^2\mu(1-\mu^2){1\over r}
{dh_0\over dr}{\partial \delta U_2\over \partial \mu}
\nonumber \\
&&+(m+2)\biggl[\case{1}{2}(m+1)^2\kappa_2{1\over r}{dh_0\over dr}
-{3m} {h_{22}\over r^2}\biggr] \delta U_0\nonumber\\
&&
-{4m(m+2)\over (m+1)^2}\pi G \bar{\rho}_0 (\delta U_0 +\delta \Phi_0)
\Biggr\}_{r=R_0}=0.
\label{4.10}
\eeqa

The operators in Eqs.~(\ref{4.9}) and (\ref{4.10}) are transformed
into the discrete matrix representation of the operator $D^i{}_j$
using the techniques discussed in Ipser and
Lindblom~\cite{ipser-lind}.  In particular we use a grid of points
$(r_j,\mu_k)$ where the $r_j$ are equally spaced in the radial
direction and the $\mu_k$ are the zeros of one of the odd-order
Legendre polynomials~\cite{notem}.  We use standard three-point
difference formulae for the derivatives in the radial direction, and
the higher-order multi-point formulae for the angular derivatives
described in Ipser and Lindblom~\cite{ipser-lind}.  Using this
discretization of the operator $D$, we find that the iteration scheme
described in Eq.~(\ref{4.8}) converges rapidly.  We begin the
iteration by setting $u^i_0=0$ and find that after about five steps
with $\Delta\lambda = -10^6R_0^2/\rho_c$ (where $\rho_c$ is the
central density) the changes in $u_n^i$ from one iteration to the next
become negligible.

Since the eigenfunction $\delta U_2$ is somewhat complicated we
present several different representations of it graphically.
Fig.~\ref{fig2} depicts the functions $\delta U_2(r,\mu_k)$ with
$\mu_k$ located at the grid points used in our integration: the roots
of $P_{19}(\mu_k)=0$ in this case.  We present in Fig.~\ref{fig3} another
representation of this $\delta U_2$ in which we graph the functions
$\delta U_2(r_k,\mu)$ for $r_k=\case{k}{5}R_0$.  This graph gives a clearer
picture of the angular structure of $\delta U_2$.  Finally we give in
Fig.~\ref{fig4} another representation of the function $\delta U_2$ in
which we decompose the angular structure of $\delta U_2$ into spherical
harmonics by defining the functions $f_k(r)$:

\bfig \centerline{\psfig{file=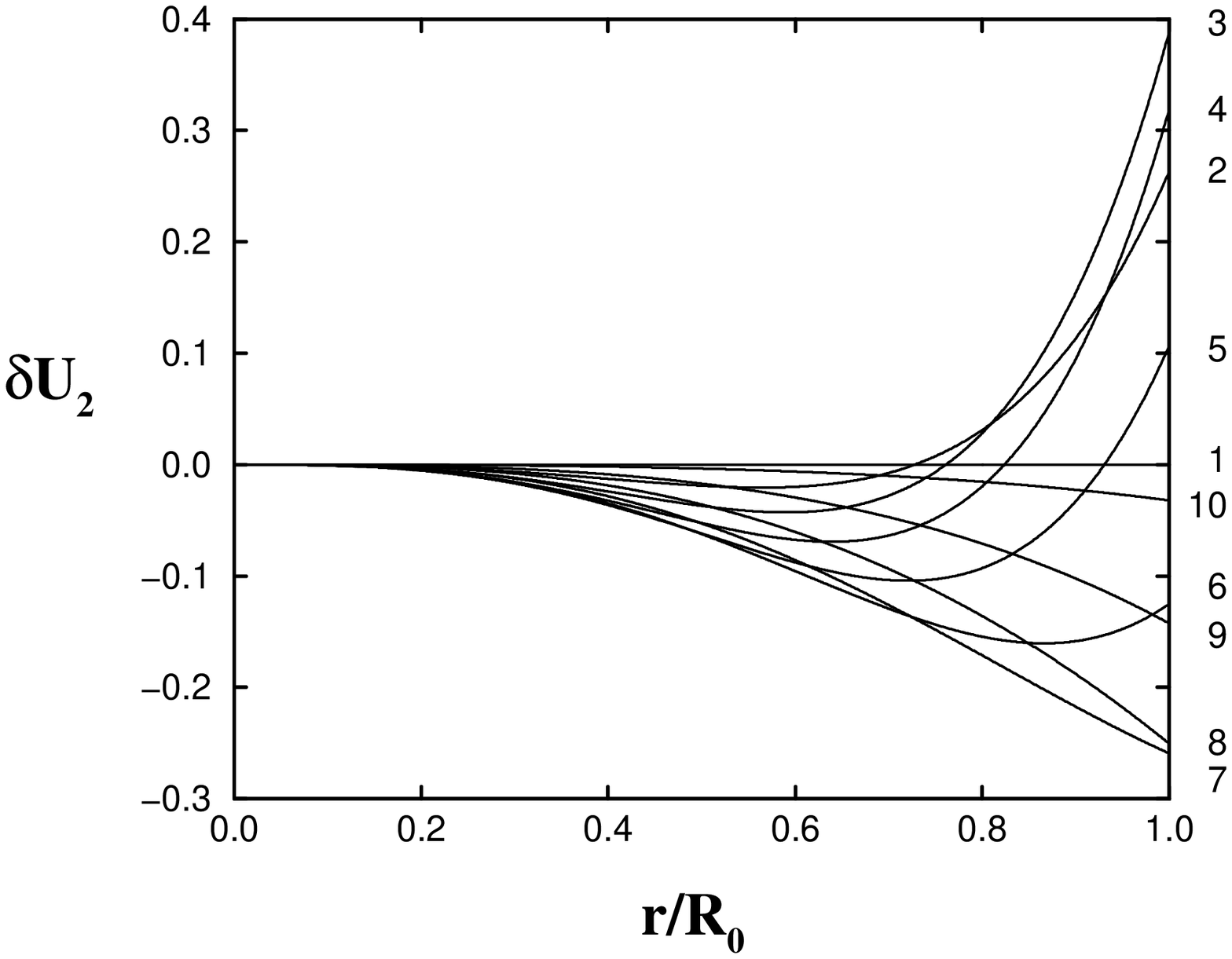,height=2.4in}} \vskip 0.3cm
\caption{Functions $\delta U_2(r,\mu_k)$ for a range of values of the
angular coordinate $\mu_k$.  The numbers along the right vertical axis
are the values of $k$.  These range sequentially from $\mu_1=0$ at the
equator of the star, to $\mu_{10}\approx 0.992$ near the rotation axis.
The equation of state is the polytrope discussed in the text.
\label{fig2}} \efig

\bfig\centerline{\psfig{file=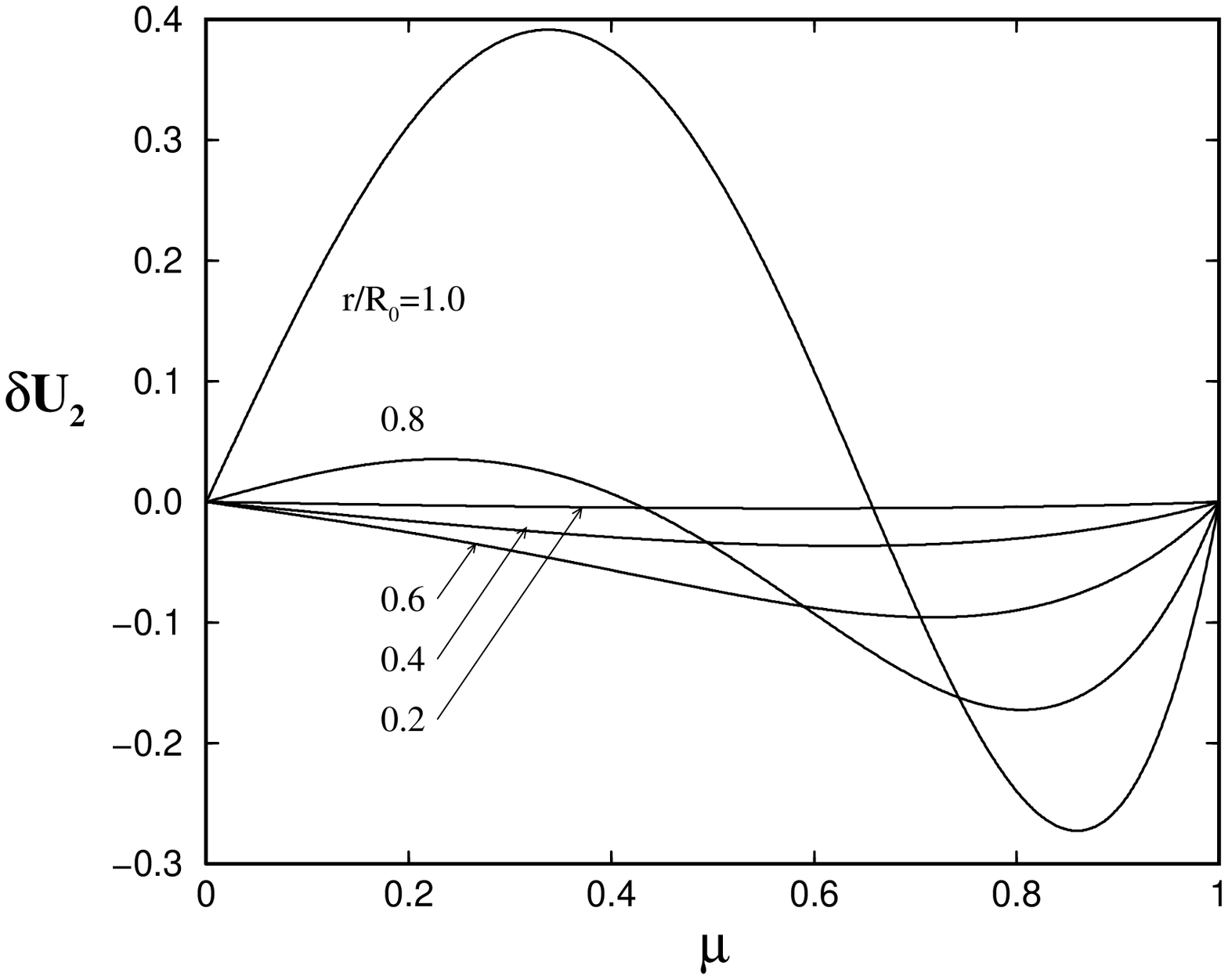,height=2.4in}}\vskip 0.3cm
\caption{Functions $\delta U_2(r_k,\mu)$ for a range of values of $r_k/R_0$.
\label{fig3}}\efig

\bfig\centerline{\psfig{file=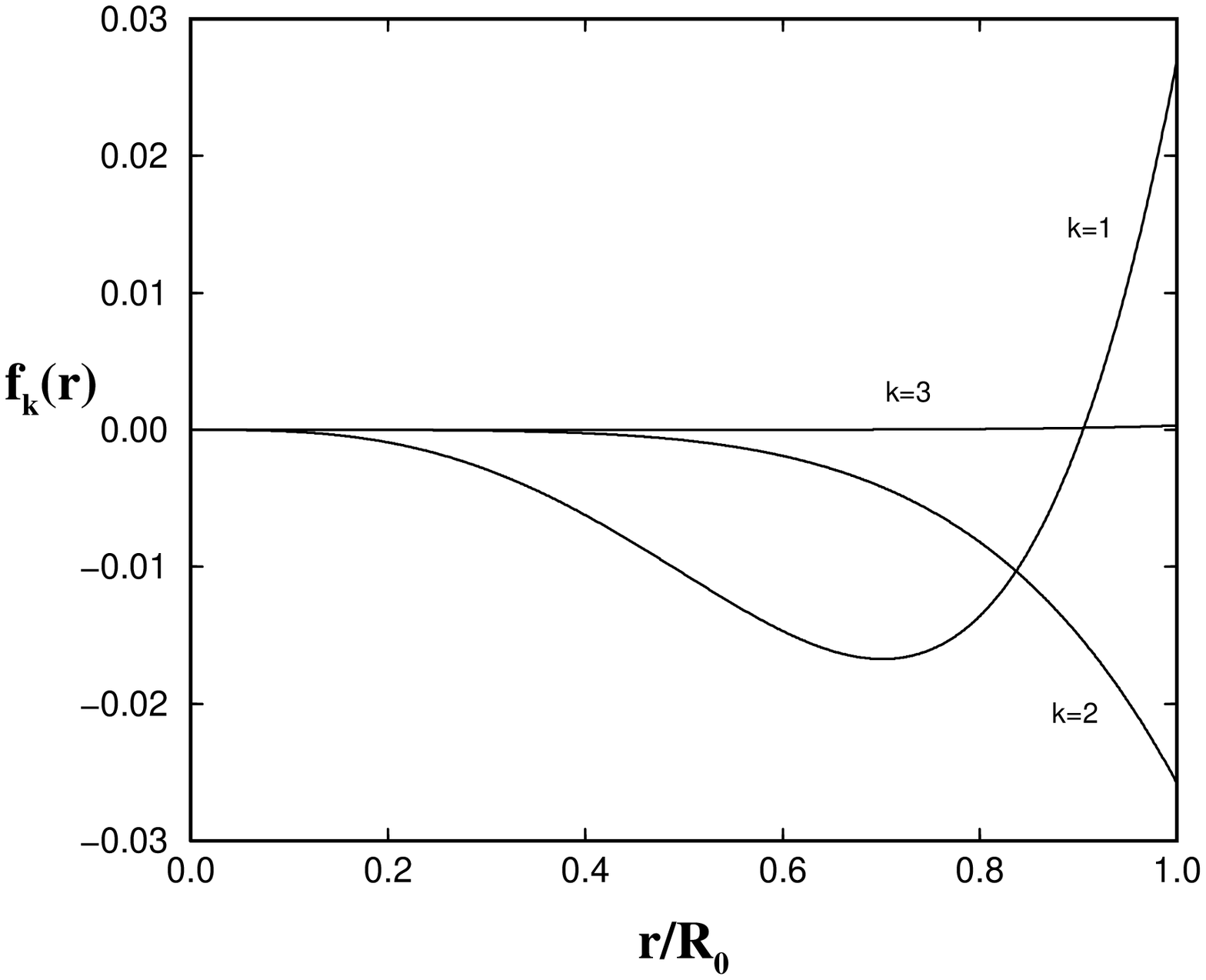,height=2.4in}}\vskip 0.3cm
\caption{Functions $f_k(r)$ that determine the spherical harmonic
decomposition of $\delta U_2$ as defined in Eq.~(\ref{4.111}).\label{fig4}}
\efig

\beq
\delta U_2(r,\mu)=\sum_{k\geq1} f_k(r)P^m_{m+2k-1}.\label{4.111}
\eeq

\noindent We find numerically that the $f_k(r)$ are negligibly small
except for the smallest few values of $k$.  In Fig.~\ref{fig4} we graph
the first three $f_k(r)$.

 To measure the degree to which $\delta U_2$ satisfies the original
differential equation, we define

\beq
\epsilon={\int|D(\delta U_2)-F|^2 r^2 drd\mu\over\int |F|^2 r^2 drd\mu}.
\label{4.11}
\eeq

\noindent We find that the value of $\epsilon$ achieved by a given
solution is approximately $\epsilon\approx (4.3/N_r)^{4}$ where $N_r$
is the number of radial grid points used in the
discretization~\cite{note001}. It is instructive to compare $\epsilon$
to the quantity

\beq
\epsilon_0 = {\int|D(\delta U_0)|^2 r^2 drd\mu\over\int |F|^2 r^2 drd\mu}.
\label{4.12}
\eeq

\noindent which measures the degree to which $\delta U_0$ is in the
kernel of $D$.  Since analytically $D(\delta U_0)=0$, the deviation of
$\epsilon_0$ from zero is a measure of the accuracy of our discrete
representation of $D$.  We find that $\epsilon_0$ is approximately
$\epsilon_0\approx (7.7/N_r)^{4}$ in our numerical solutions.  This
scaling is what is expected from the truncation errors involved in the
three-point difference formulae used to construct $D^i_j$. Since
$\epsilon<\epsilon_0$ in our numerical solutions, we see that $\delta U_2$
is a good solution to Eq.~(\ref{4.2}).


\section{Bulk Viscosity Timescales}
\label{sectionV}

One of our primary interests in evaluating these modes through
second-order in the angular velocity is our desire to obtain the
lowest-order expression for the bulk viscosity damping of these modes.
Bulk viscosity is driven by the expansion, $\delta\sigma =
\nabla_a\delta v^a$, of the fluid perturbation which can be expressed
in terms of the scalar perturbation quantities $\delta U$ and
$\delta\Phi$ as

\beq
\delta\sigma 
=-{i\over \rho}{d\rho\over dh}\Bigl[
Q^{ab}\nabla_a h \nabla_b \delta U + \kappa\Omega(\delta U +
\delta\Phi)\Bigr].
\label{5.1}
\eeq

\noindent The first term on the right side of Eq.~(\ref{5.1}) vanishes
at lowest order.  Thus the lowest-order contribution to the expansion
$\delta \sigma$ is at order $\Omega^3$.  And thus the second-order quantities
$\delta U_2$, $\kappa_2$ etc. that we have evaluated in the preceding
sections are needed to evaluate $\delta\sigma$.

Bulk viscosity causes the energy associated with a perturbation to be
dissipated according to the formula,

\beq
\left({d\tilde{E}\over dt}\right)_B
= -\int \zeta \delta \sigma\delta\sigma^* d^{\,3}x,
\label{5.2}
\eeq

\noindent where $\zeta$ is the bulk viscosity coefficient, and
$\tilde{E}$ is the energy of the perturbation as measured in the
co-rotating frame of the fluid.  The energy $\tilde{E}$ can be
expressed as an integral of the fluid perturbations:

\beq
\tilde{E} = \case{1}{2} \int \bigl(\rho \,\delta v^a\delta v^*_a
+ \delta U \delta \rho^*\bigr)d^{\,3}x.\label{5.3}
\eeq

\noindent Bulk viscosity causes the energy in a mode to decay (or
grow) exponentially with time.  We can evaluate the imaginary part of
the frequency of a mode that results from bulk-viscosity effects by
combining Eqs.~(\ref{5.2}) and (\ref{5.3}).  The result, which defines
the bulk-viscosity damping time, $\tau_B$, is given by

\beq
{1\over \tau_B} = -{1\over 2\tilde{E}}\left({d\tilde{E}\over dt}\right)_B.
\label{5.4}
\eeq

In order to evaluate $1/\tau_B$ we need
to have explicit expressions for the various terms that appear in the
integrands of Eqs.~(\ref{5.2}) and (\ref{5.3}).  The energy $\tilde{E}$,
for example, can be expressed as the integral

\beqa
\tilde{E}= {\alpha^2\pi\over 2m}&&(m+1)^3 (2m+1)!R_0^4\Omega^2
\nonumber\\&&\times
\int_0^{R_0}\rho_0(r)\left({r\over R_0}\right)^{2m+2}dr + {\cal O}(\Omega^4),
\label{5.5}
\eeqa

\noindent by performing the angular integrals indicated in
Eq.~(\ref{5.3})\cite{note01}.

In the dissipation integral, Eq.~(\ref{5.2}), an explicit expression
for the bulk viscosity $\zeta$ is needed.  In standard neutron-star
matter the dominant form of bulk viscosity is due to the emission of
neutrinos via the modified URCA process~\cite{sawyer}.  An approximate
expression for this form of the bulk-viscosity coefficient
is~\cite{ipser-lindblom.ii}

\beq
\zeta = 6.0\times 10^{-59}{\rho^2T^6\over \kappa^2\Omega^2},
\label{5.7}
\eeq

\noindent where all quantities are expressed in cgs units.  For the
case of the classical $r$-modes the expansion $\delta \sigma$ that
appears in Eq.~(\ref{5.2}) can be expressed explicitly in terms of the 
potentials $\delta U_0$, $\delta\Phi_0$, and $\delta U_2$:

\beqa
\delta &&\sigma = {i\over \rho_0}\left({d\rho\over dh}\right)_0
{R_0^2\Omega^3\over \pi G \bar{\rho}_0} {m+1\over 2m(m+2)}\nonumber\\
&&\times\Biggl\{
{m(m+1)\over r}{dh_0\over dr}\delta U_2
+[1-(m+1)^2\mu^2]{dh_0\over dr}{\partial\delta U_2\over \partial r}
\nonumber \\
&&\quad
-(m+1)^2 {\mu(1-\mu^2)\over r}{dh_0\over dr}{\partial\delta U_2\over 
\partial \mu}
-3m(m+2)h_{22}{\delta U_0\over r^2}\nonumber \\
&&\quad
+\kappa_2 {(m+2)(m+1)^2\over 2r}{dh_0\over dr}\delta U_0\nonumber \\
&&\quad
-4\pi G \bar{\rho}_0 {m(m+2)\over (m+1)^2}(\delta U_0+\delta\Phi_0)
\Biggr\}+ {\cal O}(\Omega^5).
\label{5.8}
\eeqa

It will be of some interest to evaluate the accuracy of some
previously published approximations for the bulk-viscosity timescale.
One of these~\cite{lom} is based on an approximation
$\delta\sigma\approx\delta s$ for the expansion of the mode, where

\beqa
\delta s &&\equiv -i\kappa\Omega {\delta\rho\over \rho}\nonumber \\
&&= -{2i\over m+1}{1\over \rho_0}\left({d\rho\over dh}\right)_0
\bigl(\delta U_0+\delta \Phi_0\bigr)R_0^2\Omega^3 + {\cal O}(\Omega^5).
\nonumber \\
\label{5.9}
\eeqa

\noindent We note that $\delta s$ is just the last term in the
expression given in Eq.~(\ref{5.8}) for $\delta \sigma$.  It is the
only term in Eq.~(\ref{5.8}) that depends only on the lowest-order
perturbation quantities: the others depend on higher-order
corrections through $\delta U_2$, $\kappa_2$ or $h_{22}$.  We define
the approximate bulk-viscosity timescale $\tau_{\tilde{B}}$ in analogy
with Eq.~(\ref{5.4}) by replacing $\delta \sigma$ with $\delta s$ in
Eq.~(\ref{5.2}).

The bulk-viscosity contribution to the imaginary
part of the frequency, $1/\tau_B$, is proportional to $\Omega^2$.  This
follows from Eqs.~(\ref{5.2}) and (\ref{5.4}) because $\tilde{E}$
scales as $\Omega^2$ from Eq.~(\ref{5.5}), $\zeta$ as $\Omega^{-2}$
from Eq.~(\ref{5.7}), and $\delta\sigma$ as $\Omega^3$ from
Eq.~(\ref{5.8})~\cite{note1}.  The bulk-viscosity damping time,
$1/\tau_B$, also scales with temperature as $T^6$.  Thus it is
convenient to define $\tilde{\tau}_B$: the bulk viscosity timescale
evaluated at $\Omega^2 = \pi G \bar{\rho}_0$ and $T=10^9K$,

\beq
{1\over \tau_B} = {1\over \tilde{\tau}_B} \left({T\over 10^9{\rm K}}
\right)^6\left({\Omega^2\over \pi G \bar{\rho}_0}\right)
+{\cal O}(\Omega^4).\label{5.10}
\eeq

\noindent We have evaluated $\tilde{\tau}_B$ numerically for the $m=2$
$r$-mode (the one most unstable to gravitational radiation) of a
$1.4M_\odot$ stellar model with the polytropic equation of state
discussed in Sec.~\ref{sectionI}, and find $\tilde{\tau}_B=
2.01\times 10^{11}$s~\cite{note2}.  For comparison, we have
re-evaluated the approximate timescale $\tau_{\tilde{B}}$ described
above and find $\tilde{\tau}_{\tilde{B}}=
7.04\times10^9$s~\cite{note3}. 

The most interesting application of these dissipative timescales is to
use them to evaluate the stability of rotating neutron stars to the
gravitational radiation driven instability in the
$r$-modes~\cite{lom}.  The imaginary part of the frequency of the
$r$-modes, $1/\tau$, includes contributions from gravitational
radiation $\tilde{\tau}_{GR}$ in addition to shear $\tilde{\tau}_S$
and bulk $\tilde{\tau}_B$ viscosity effects.  The general expression
for $\tau(\Omega,T)$, a function of the temperature $T$ and angular
velocity $\Omega$ of the star, is given by

\beqa
{1\over \tau(\Omega,T)}= &&{1\over \tilde{\tau}_{GR}}
\left({\Omega^2\over \pi G \bar{\rho}}\right)^3
+ {1\over\tilde{\tau}_S}\left({10^9{\rm K}\over T}\right)^2\cr
&&\qquad+ {1\over\tilde{\tau}_B}\left({T\over 10^9{\rm K}}\right)^6
\left({\Omega^2\over \pi G \bar{\rho}}\right).\label{5.11}
\eeqa

\noindent The bulk viscosity timescale $\tilde{\tau}_B=2.01\times
10^{11}$s has been evaluated in this paper, while the gravitational
radiation timescale, $\tilde{\tau}_{GR}=-3.26$s, and the shear
viscosity timescale $\tilde{\tau}_S=2.52\times 10^8$s, were obtained
by Lindblom, Owen and Morsink~\cite{lom} for the polytropic stellar
model discussed in Sec.~\ref{sectionI}.  It is interesting to
determine from this expression the critical angular velocity
$\Omega_c$:

\beq
{1\over \tau(\Omega_c,T)}=0.
\label{5.12}
\eeq

\noindent For stars at a given temperature, those with
$\Omega>\Omega_c$ are unstable to the gravitational radiation driven
instability in the $r$-modes, while those rotating more slowly are
stable.  Fig.~\ref{fig5} depicts $\Omega_c$ for a range of
temperatures relevant for hot young neutron stars.  For stars cooler
than about $10^9$K, superfluid effects change the dissipation
processes completely and the analysis presented here is no longer
relevant.  The dashed curve in Fig.~\ref{fig5} represents the critical
angular velocities computed using the approximate bulk viscosity
damping time $\tilde{\tau}_{\tilde B}=7.05\times 10^9$s instead of the
exact value.  We see that even though $\tilde{\tau}_{B}\approx 29
\tilde{\tau}_{\tilde B}$, the qualitative shape of the $\Omega_c$
curve is not affected.  The minimum of the $\Omega_c(T)$ curve
depicted in Fig.~\ref{fig5} occurs at $\min\Omega_c = .0301\sqrt{\pi G
\bar{\rho_o}}$ which is 4.51\% of the approximate maximum angular
velocity $\case{2}{3}\sqrt{\pi G \bar{\rho_o}}$. 

\bfig
\centerline{\psfig{file=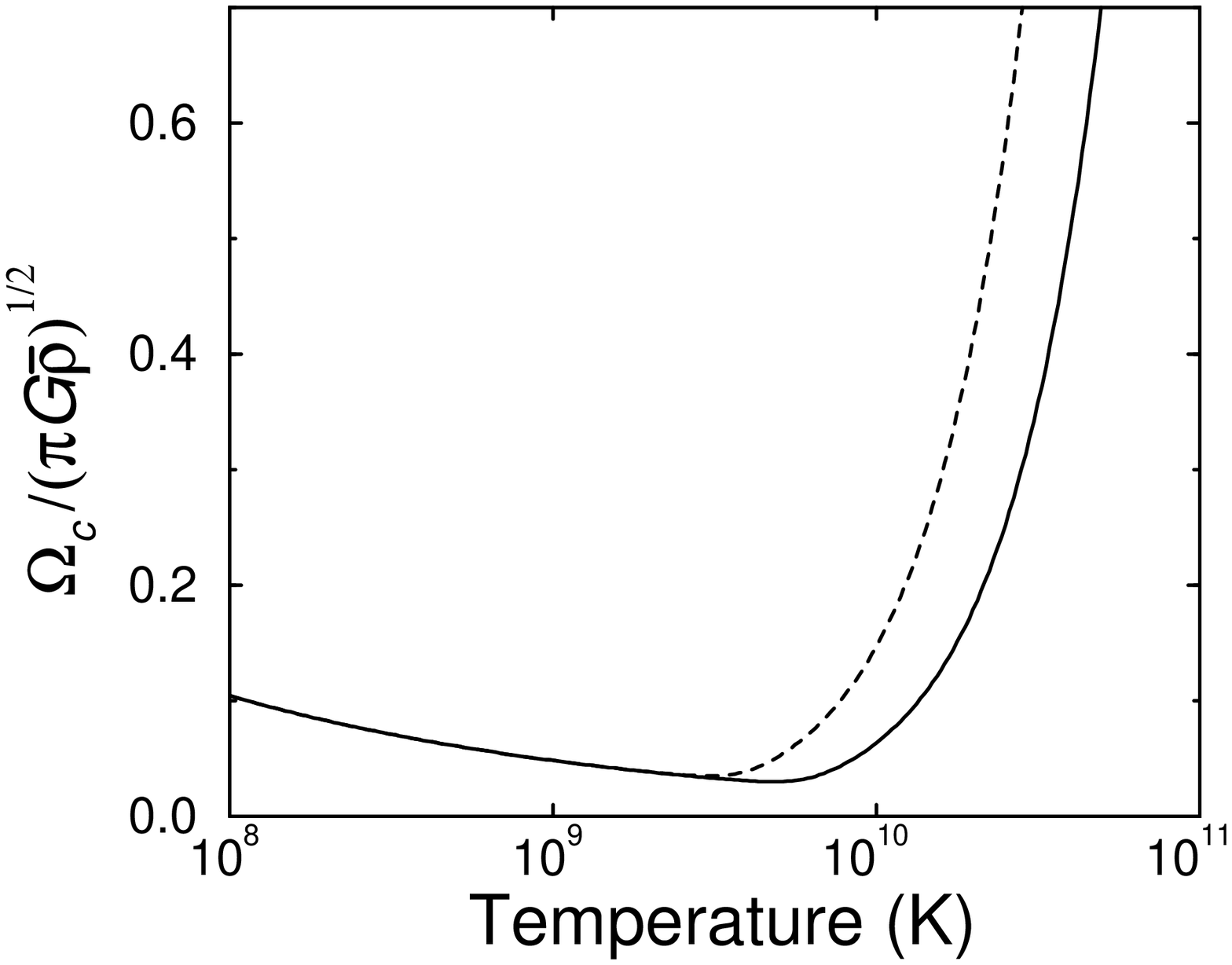,height=2.4in}}\vskip 0.3cm
\caption{Critical angular velocities for the $r$-mode instability
as a function of temperature.  The dashed curve gives the
critical angular velocities based on the approximate bulk
viscosity damping time $\tau_{\tilde{B}}$.\label{fig5}}
\efig

Stars composed of strange quark matter are subject to a different form of
bulk viscosity caused by weak interactions that transform $d$ quarks
to $s$ quarks.  The bulk viscosity coefficient that results from this
process is given approximately by~\cite{madsen1}

\beq
\zeta = 3.2\times 10^3 {\rho T^2\over \kappa^2\Omega^2} 
\left({m_s\over 100 {\rm Mev}}\right)^4,
\label{5.13}
\eeq

\noindent where all quantities (except $m_s$, the mass of the $s$ quark)
are given in cgs units.  We have used this form of the bulk viscosity
to estimate the damping time of the $r$-modes in strange stars.  We find
that in strange stars this damping time scales as

\beq
{1\over \tau_B} = {1\over \tilde{\tau}_B} \left({T\over 10^9{\rm K}}
\right)^2\left({\Omega^2\over \pi G \bar{\rho}_0}\right)
\left({m_s\over 100 {\rm Mev}}\right)^4.
\label{5.14}
\eeq

\noindent Using the polytropic stellar model described in
Sec.~\ref{sectionI}, we have computed the bulk-viscosity damping time
of the $r$-mode to be $\tilde{\tau}_B=0.886$s for strange stars.  This
value is smaller (by about a factor of 7) than that found by
Madsen~\cite{madsen2}, who used a very rough estimate of the
bulk-viscosity damping time~\cite{KS} for the $r$-modes.  We also note
that our expression for $\tau_B$ does not scale with angular velocity
in the same way as Madsen's.  Nevertheless, our calculations confirm
Madsen's prediction that bulk viscosity completely suppresses the
$r$-mode instability in hot strange stars.  Using our expression for
${\tau}_B$ we estimate that the $r$-mode instability is suppressed in
all strange stars with $T\gtrsim 5\times 10^8$K.

\acknowledgments We especially thank Y.~Levin for helping to
stimulate our interest in finding the techniques needed to evaluate
accurately the bulk-viscosity coupling to the $r$-modes.  We also
thank J.~Creighton, S.~Detweiler, J.~Ipser, N.~Stergioulas, K.~Thorne,
and R.~Wagoner for helpful conversations concerning this work.  This
research was supported by NSF grants AST-9417371 and PHY-9796079, and
by NASA grant NAG5-4093.

\appendix
\section*{Numerical Relaxation}
\label{appendix}

The operator $D$ defined in Eq.~(\ref{4.1}) is symmetric, in the sense
that

\beq
\int g^* D(f) d^{\,3}x = \left[\int f^* D(g) d^{\,3}x\right]^*,\label{a1}
\eeq

\noindent for arbitrary functions $f$ and $g$ in any stellar model
where $\rho_0=0$ on the surface.  Thus the discrete representation of
the operator $D_{ij}$ will be a Hermitian matrix, and consequently
$D^i{}_j$ will have a complete set of eigenvectors.  Let $e^i_\alpha$
denote the eigenvector corresponding to eigenvalue $d_\alpha$:
$D^i{}_je^j_\alpha = d_\alpha e^i_\alpha$.  Since these eigenvectors
form a complete set, we can express any vector as a linear combination
of them.  Thus we take $F^i = \sum_\alpha F^\alpha e^i_\alpha$, $u^i_n
= \sum_\alpha u^\alpha_n e^i_\alpha$, etc.  The numerical relaxation scheme
indicated in Eq.~(\ref{4.8}) can be re-expressed therefore in the
eigenvector basis as:

\beq
u^\alpha_{n+1} = 
{\Delta\lambda F^\alpha - u^\alpha_n\over
\Delta\lambda d_\alpha -1}.
\label{a2}
\eeq

\noindent The role of the projection operator $P^i{}_j$ is merely
to remove from Eq.~(\ref{a2}) the component corresponding to the zero
eigenvalue.  The recurrence relation Eq.~(\ref{a2}) can be solved
analytically:

\beq
u^\alpha_{n+1} = x_\alpha \Delta \lambda F^\alpha \sum_{k=0}^n(-x_\alpha)^k,
\label{a3}
\eeq

\noindent where

\beq
x_\alpha = {1\over \Delta\lambda d_\alpha -1}.\label{a4}
\eeq

\noindent  This series converges as long as $|x_\alpha|<1$.  Since
the projection operator has eliminated the one equation where $d_\alpha=0$
it is easy to choose $\Delta\lambda$ so that $|x_\alpha|<1$ for all $\alpha$,
e.g., by taking $\Delta\lambda$ sufficiently large.
Thus, the sequence $u^\alpha_{n+1}$ converges to

\beq
\lim_{n\rightarrow\infty}u^\alpha_{n} 
= {x_\alpha\Delta \lambda F^\alpha\over1+x_\alpha}={F^\alpha\over d_\alpha}.
\label{a5}
\eeq

\noindent Thus the implicit relaxation scheme converges to the desired
solution to Eq.~(\ref{4.2}).

In contrast to the implicit relaxation method defined in
Eq.~(\ref{4.8}), the analogous explicit relaxation method does not
converge at all for this problem.  An analysis similar to that carried
out above reveals that the criterion for convergence of the explicit
scheme is that $|\Delta \lambda d_\alpha + 1|<1$.  Clearly this can
only hold for operators $D$ where the eigenvalues all have the same
sign (as is the case when $D$ is elliptic) and only when $\Delta
\lambda$ has the correct sign.  Our limited experience
with hyperbolic operators $D$ is that their eigenvalues have both
signs.  Consequently it is not surprising that our attempts at
explicit numerical relaxation fail in this case.



\begin{references}
\bibitem{note0} Throughout this paper we use ``second order'' to
  describe terms in the expansion of a quantity in powers of the
  angular velocity which are two powers in the angular velocity higher
  than the lowest-order terms.  For example, if the lowest-order
  velocity perturbation were normalized so that it was linear in the
  angular velocity, then the second-order terms would be proportional
  to the angular velocity cubed.
\bibitem{notem} We have also written a second totally independent code
that uses a uniformly spaced angular grid, and standard three-point
angular differencing.  We find the results of the two independent
codes to be completely equivalent.  However, the code based on the
uniformly spaced angular grid requires many more angular grid points
than the code described more fully in the text to achieve the same
accuracy.
\bibitem{note001} The value of $\epsilon$ is essentially independent
of the number of angular grid points $N_a$ because the angular
dependence of the function $\delta U_2$ is determined by only a few
terms when expessed as the series in Eq.~(\ref{4.111}).  For such
functions the angular differencing method discussed in Ipser and
Lindblom~\cite{ipser-lind} is essentially exact.
\bibitem{note01}  We note that the normalization of these modes used
here (as defined in Eq. \ref{3.1}) differs from that used in Lindblom,
Owen, and Morsink~\cite{lom}.  The normalization parameter $\alpha$ used
here is related to the one used there (call it $\alpha'$) by
\beqa
(\alpha')^2 = {\alpha^2\pi\over m}(m+1)^3(2m+1)!\nonumber
\eeqa
\bibitem{note1} We note that Andersson, Kokkotas, and
Schutz~\cite{a.k.s} report that $1/\tau_B$ scales as $\Omega^{1.77}$
instead of $\Omega^2$ as demanded by the small angular velocity
expansion used here.
\bibitem{note2} We note that ${\tau}_{B}$ found here is about a factor
of four larger than that reported by Andersson, Kokkotas, and
Schutz~\cite{a.k.s}, when we compare values at $T=10^9$K and
$\Omega^2=\pi G\bar{\rho}_0$.  Since they use a different scaling of
$\tau_B$ with $\Omega$, the discrepancy between our values is angular
velocity dependent: the ratio of our timescale to their's is
$3.7 \bigl(\sqrt{\pi G \bar{\rho}_0}/\Omega\bigr)^{.23}$.
\bibitem{note3} We note that this value of the approximate timescale
$\tilde{\tau}_{\tilde{B}}$ is about ten times the value published by
Lindblom, Owen, and Morsink~\cite{lom}.  This discrepancy is due to an
error in Eq.~(4) of that paper: the right side of that equation should
be multiplied by an additional overall factor of
$(2l+1)/[l(l+1)\sqrt{2l+3}]$.  Consequently the numerical bulk
viscosity damping time estimate given there must be multiplied by
$l^2(l+1)^2(2l+3)/(2l+1)^2$, or about 10.08 for the $l=2$ case.
\end{references}
\end{document}